\pdfoutput=1
\documentclass[12pt]{iopart}

\usepackage{pdflscape} 
\usepackage{amsfonts}
\usepackage{amssymb}
\usepackage{graphicx}
\usepackage{epstopdf}
\usepackage{verbatim}

\usepackage{wasysym}
\usepackage{setstack}
\usepackage[thickspace,amssymb]{SIunits}
\usepackage[utf8]{inputenc}
\usepackage{csquotes}

\usepackage{harvard}
\begin{document}

\title[]{Secondary radiation measurements for particle therapy applications: nuclear fragmentation produced by $^{4}$He ion beams in a PMMA target}

\author{M.~Marafini$^{a,b}$, R.~Paramatti$^{a,c}$, D.~Pinci$^{a}$, G.~Battistoni$^{d}$, F.~Collamati$^{e}$, E.~De Lucia$^{e}$, R.~Faccini$^{a,c}$, P.~M.~Frallicciardi$^{f}$, C.~Mancini-Terracciano$^{a,c}$, I.~Mattei$^{d}$, S.~Muraro$^{d}$, L.~Piersanti$^{a,g}$, M.~Rovituso$^{h}$, A.~Rucinski$^{a,g}$, A.~Russomando$^{a,c,i}$, A.~Sarti$^{b,d,g}$, A.~Sciubba$^{a,b,g}$, E.~Solfaroli~Camillocci$^{a,c}$, M.~Toppi$^{e}$, G.~Traini$^{a,c}$, C.~Voena$^{a}$, V.~Patera$^{a,b,g}$}

\address{$^a$ INFN - Sezione di Roma, Italy}
\address{$^b$ Museo Storico della Fisica e Centro Studi e Ricerche ``E.~Fermi'', Roma, Italy}
\address{$^c$ Dipartimento di Fisica, Sapienza Universit\`a di Roma, Italy}
\address{$^d$ INFN - Sezione di Milano, Italy}
\address{$^e$ Laboratori Nazionali di Frascati dell'INFN, Frascati, Italy} 
\address{$^f$ Istituto di ricerche cliniche Ecomedica, Empoli, Italy}
\address{$^g$ Dipartimento di Scienze di Base e Applicate per Ingegneria, Sapienza Universit\`a di Roma, Italy}
\address{$^h$ GSI Helmholtzzentrum f$\ddot{u}$r Schwerionenforschung, Darmstadt, Germany}
\address{$^i$ Center for Life Nano Science@Sapienza, Istituto Italiano di Tecnologia, Roma, Italy}

\eads{(corresponding author) \mailto{riccardo.paramatti@roma1.infn.it}}



\begin{abstract}

Nowadays there is a growing interest in Particle Therapy treatments exploiting light ion beams against tumors due to their enhanced Relative Biological Effectiveness and high space selectivity. In particular promising results are obtained by the use of $^4$He projectiles. Unlike the treatments performed using protons, the beam ions can undergo a fragmentation process when interacting with the atomic nuclei in the patient body. In this paper the results of measurements performed at the Heidelberg Ion-Beam Therapy center are reported. For the first time the absolute fluxes and the energy spectra of the fragments - protons, deuterons, and tritons - produced by $^4$He ion beams of $102$, $125$ and $145\ \mega\electronvolt/u$ energies on a poly-methyl methacrylate target were evaluated at different angles.
The obtained results are particularly relevant in view of the necessary optimization and review of the Treatment Planning Software being developed for clinical use of $^4$He beams in clinical routine and the relative bench-marking of Monte Carlo algorithm predictions.
\end{abstract}


 \maketitle

\section{Introduction}

The efficacy of the treatment of radio-resistant tumors with Particle Therapy (PT) is nowadays well established, and the number of centers that can carry out a PT treatment is steadily increasing. The advantage of using hadron beams with respect to the conventional radio therapy is related to the mechanism of energy loss in matter, characterized by a loss of a small fraction of energy in the first part of the hadron path within the patient body with a following release of almost all the hadron energy in the very small region where the energy loss per distance traveled reaches its maximum, called Bragg Peak (BP). 

In the interaction with the atomic nuclei of the patient body, the ions can fragment in nuclei with lower atomic number Z which penetrate more in depth causing an energy loss tail beyond the BP region. A precise knowledge of the ion-target cross section for different type of fragments is therefore fundamental to estimate the additional dose absorbed by the healthy tissues and organs at risk surrounding the tumor.
Detailed measurements were performed in the past on the fragmentation of heavy ion beams as, for example, on carbon~\cite{Haettner2013,Gunzert-Marx2008}.

In last years, there is a growing interest in the usage of beams of $^4$He ions for their high Linear Energy Transfer (LET) and Relative Biological Effectiveness (RBE) characteristics~\cite{tommasino2016,Kramer2016,Durante2016}.

In this work we present detailed measurements of the production of protons ($p$), deuterons ($d$) and tritons ($t$) by an $^4$He ion beam impinging on a Poly Methyl Methacrylate (PMMA) target. Data were collected at the Heidelberg Ion-Beam Therapy center (HIT), in Germany, with $^4$He beams of $102$, $125$ and $145\ \mega\electronvolt/u$ kinetic energy. The absolute fluxes and the angular and energy spectra of secondary fragments were studied as a function of the beam energy.

A precise measurement of the projectile fragmentation process cross section is essential to correctly evaluate the corresponding dose released to the patient and take it into account in the Treatment Planning Software (TPS) \cite{Kramer2000}.

A similar work~\cite{Rovituso2016} presents the loss of $^4$He beams on water and PMMA at different depths and the double differential fragment yields with water targets. 

In this paper the experimental setup is described in Section~\ref{sec:setup}, the strategy of the measurement and the particle identification are discussed in details in Section~\ref{sec:misure}. Section~\ref{sec:MC} summarizes the correction factor definition and computation, like geometrical acceptance and detector efficiency, to be applied to the fragment flux measurements. Finally the cross section results are reported in Section~\ref{sec:exp}. 

\clearpage
\section{Experimental setup}
\label{sec:setup}

Beams of $^4$He ions of different energies relevant for PT have been used to irradiate a PMMA target in order to study the projectile fragmentation at different angles. Such measurements were part of a complex experiment, exploited with a multi-purpose detector setup, aiming for the characterization of the interactions of different ion species of several energies with the purpose of studying the secondary particle emission trough the measurement of prompt photons, charged and $\beta^+$-emitter fragments. More details can be found in~\cite{Mattei2016,Rucinski2016,Toppi2016}.

\begin{figure}[!ht]
\begin{center}
\includegraphics [width = 0.9 \textwidth] {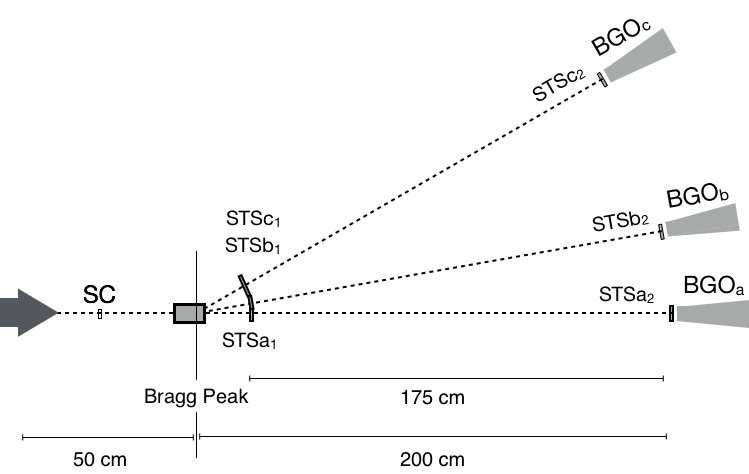}
\caption{\small{Top view of the experimental setup. The beam runs form the left and crosses the Start Counter (SC) before impinging on the PMMA target. On the right three measurement arms: two Time of Flight detectors (STS) and a crystal calorimeter (BGO) per arm.}}
\label{fig:Schema}
\end{center}
\end{figure}

A simplified view of the HIT experimental set-up is sketched in Fig.~\ref{fig:Schema} showing only the detector and geometrical configuration that is relevant for the forward fragmentation studies. The origin of the reference frame was placed in the Bragg Peak position, evaluated for each beam energy by means of a dedicated Monte Carlo software, shortly before the PMMA exit face. The beam, running from left to right in the figure, came out from the vacuum pipe about 50 $\centi\meter$  upstream the PMMA target. 
It had a Full Width at Half Maximum size dependent on the energy and ranging from 6.9 mm to 9.3 mm in the transverse plane. To detect the incoming primary particles, a $2\ \milli\meter$ thick plastic scintillator (EJ$200$), readout by two Hamamatsu H$6524$ photo-multiplier tubes (PMTs) and referred as Start Counter (SC) in the following, was placed at about $37\ \centi\meter$ upstream from the PMMA. The time resolution of the SC tubes was measured to be about $250\ \pico\second$.

In order to keep the BP position close to the end of PMMA, its thickness (t$_{\rm PMMA}$) was adjusted accordingly to the specific range of each beam, computed with a FLUKA MC simulation~\cite{Ferrari2005,Battistoni2007,Battistoni2015}. The beam energies and the corresponding values for the BP depth and t$_{\rm PMMA}$ are reported in Table~1. 
The PMMA face dimensions were $5.0 \times 5.0\ \centi\meter^2$. 

\begin{table}[h]
\begin{center}
\label{tab:datatake}
\caption{Bragg Peak depth (with respect to the ion entrance point) and thickness of the PMMA target for data taking at different beam energies.}
\begin{tabular}{|c|c|c|} 
\hline
$E_{beam}$ & $BP_{Depth}$ & $t_{\rm PMMA}$\\
$(\mega\electronvolt/u) $ & $(\centi\meter)$ & $(\centi\meter)$\\
\hline
102.34 &  6.68 &  7.65 \\ 
124.78 &  9.68 & 10.00 \\ 
144.63 & 12.63 & 12.65 \\ 
\hline
\end{tabular}
\end{center}
\end{table}

Three identical arms ($a, b$ and $c$ in Fig. \ref{fig:Schema}) were assembled and placed downstream from the PMMA. Each arm was instrumented with:
\begin{itemize}
\item two thin scintillators, STS$_1$ ($4.0 \times 4.0 \times 0.2\ \centi\meter^3$) and STS$_2$ ($4.0 \times 4.0 \times 1.0\ \centi\meter^3$), read by a H$10580$ PMT, placed $175\ \centi\meter$ apart from each other, to measure the fragment Time of Flight (ToF); 

\item a matrix of $2 \times 2$ Bi$_4$Ge$_3$O$_{12}$ (BGO) crystals readout by an EMI-$9814$B PMT to measure the energy deposited by the fragments. Each crystal has a trunk pyramid shape with a front face of $2.3 \times 2.3\ \centi\meter^2$, a rear face of $3.1 \times 3.1\ \centi\meter^2$ and a length of $20\ \centi\meter$. The matrices were placed straight after the STS$_2$ at a distance of about $2\ \meter$ from the origin of the reference frame (see Fig.~\ref{fig:Schema})
\end{itemize}

Analog and discriminated signals were acquired by a VME system instrumented by a $19$-bit multi-hit TDC (CAEN V1190B) with a time resolution of 100 ps, a $12$-bit QDC (CAEN V792N) with a charge resolution of 0.1 pC and a $32$-bit scaler (CAEN V560N) able to operate up to a maximum input rate of $100 \ \mega\hertz$. The effective dynamic range of the QCD was increased by splitting in three the response of the BGO-PMT and sending it, via different attenuators, to three channels of the QDC. The time gate chosen for the BGO signal acquisition was $1\ \micro\second$.

The performance of all these detectors were studied and optimized in laboratory by means of cosmic rays and radioactive sources:
\begin{itemize}
\item[-] the time resolution of the STS$_1$ was measured to be 250 $\pico \second$;
\item[-] given their larger thickness, the STS$_2$ have a slightly better time resolution of about $200\ \pico \second$ and are able to measure an energy release with a precision of about 10\%;
\item[-] the energy resolution of a single BGO crystal was found to have a stochastic contribution of about 20\%/$\sqrt{E[\mega\electronvolt]}$ with an effective constant term always below $1\ \mega\electronvolt$. However, since the crystal responses were not equalized among them and the lateral containment was not perfect, the effective energy resolution of the matrices was about ten times larger. The time resolution of the BGO was measured to be about $700\ \pico\second$.
\end{itemize}

By placing these arms at different positions, it was possible to investigate the fragment fluxes in five angular configurations: $0\degree$ and $5\degree$ (arm $a$), $10\degree$ and $15\degree$ (arm $b$) and $30\degree$ (arm $c$). The measurement at $30\degree$ has been performed twice to test the reproducibility of the results. 

\subsection{Trigger and rates}
\label{sec:rates}

The trigger signal was provided by the coincidence, within a 80 $\nano\second$ time window, of the logical OR of the two SC photo-multipliers (SCOR) and at least one of the BGO matrix discriminated signals. 

The threshold used to discriminate the BGO signal was equivalent to an energy release in the crystal matrix of about $20\ \mega\electronvolt$. The output rate of the discriminator for the BGO in arm $a$ was down-scaled to not saturate the DAQ capabilities and not suppress the event rate at the other angles.

At each event, the trigger signal was vetoed during data conversion giving rise to a dead time ({\it DAQ dead time} or $\delta_{\rm DT}$ in the following) of the order to few hundreds of $\micro\second$.

The beam rate ranged from about 300 $\kilo\hertz$ up to 3 $\mega\hertz$, while the trigger and DAQ rates ranged from about 300~$\hertz$ to 2~$\kilo\hertz$ (being the DAQ rate limited to a maximum of 6~$\kilo\hertz$). The fraction of acquired events triggered by the BGO was about the 85\% of the total. In order to estimate the effect of $\delta_{\rm DT}$ on the total efficiency, the vetoed and the un-vetoed trigger rates were acquired by the scaler. The value of $\delta_{\rm DT}$ is therefore defined as the number of vetoed triggers over the number of un-vetoed triggers and it depends on the beam rate. 

The total number of primary ions $N_{He}$ has been computed using the information provided by the SC. Since the beam rate was well below the maximum operating frequency of the scaler, $N_{He}$ had to be corrected only for the effect of the dead time introduced by the width of the SC discriminated signal (100~$\nano\second$). This correction $C_{He}$ has been evaluated by means of a dedicated toy Monte Carlo that reproduces the temporal beam structure of each data taking. The value of $C_{He}$ ranges from $1.2$ to $1.6$, depending on the beam rate. A detailed description of how the correction factors and their statistical and systematic uncertainties have been computed can be found in~\cite{Mattei2016}.

\clearpage
\section{Measurements strategy}
\label{sec:misure}

The study of the interactions of the $^4$He ions with a PMMA target and of the fragmentation that can occur resulting in the production of protons, deuterons and tritons, proceeds through the correct fragment identification and kinetic energy measurements. The adopted strategies and the related performances are outlined below.

\subsection{Particle Identification}
\label{sec:pid}

In order to separate the different populations of projectile fragments, several observables were measured event by event by each arm:
\begin{itemize}
\item[-]
$E$: the deposited energy in the BGO matrix.
\item[-]
$\Delta E$: the energy released in the STS$_2$ in front of the BGO.
\item[-]
$ToF$: the Time of Flight between the two scintillators STS$_1$ and STS$_2$.
\item[-]
$\Delta t_{BGO-SC}$: the time interval between the passage of the beam ion in the Start Counter and the signal detected in the BGO matrix.
\end{itemize}

\begin{figure}[!ht]
\begin{center}
\includegraphics [width = 0.48 \textwidth] {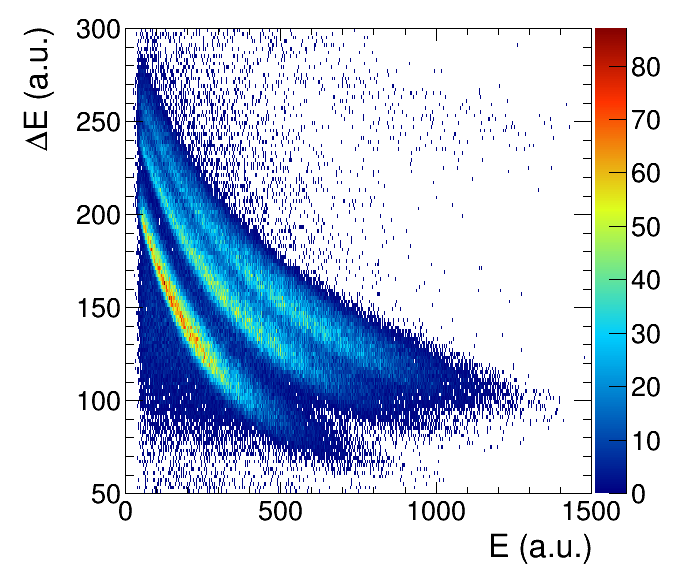}
\includegraphics [width = 0.48 \textwidth] {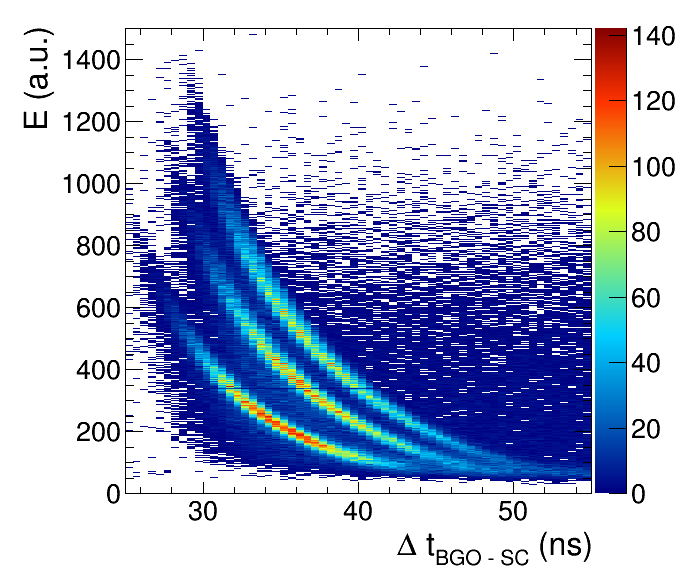}
\includegraphics [width = 0.48 \textwidth] {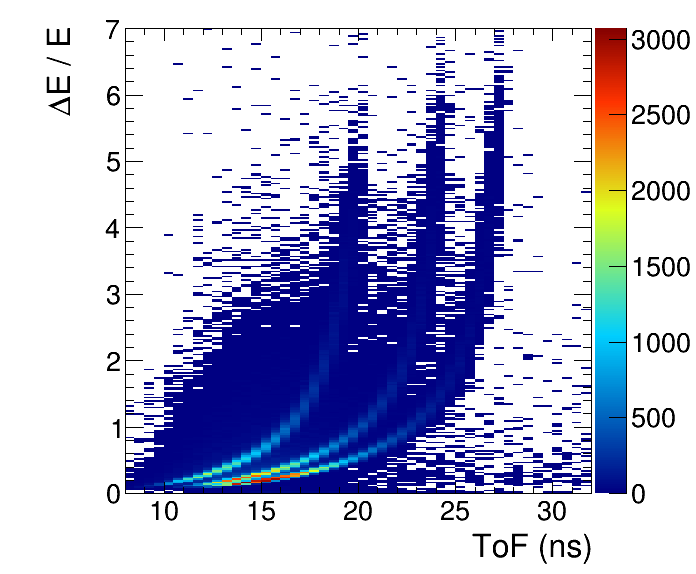}
\includegraphics [width = 0.48 \textwidth] {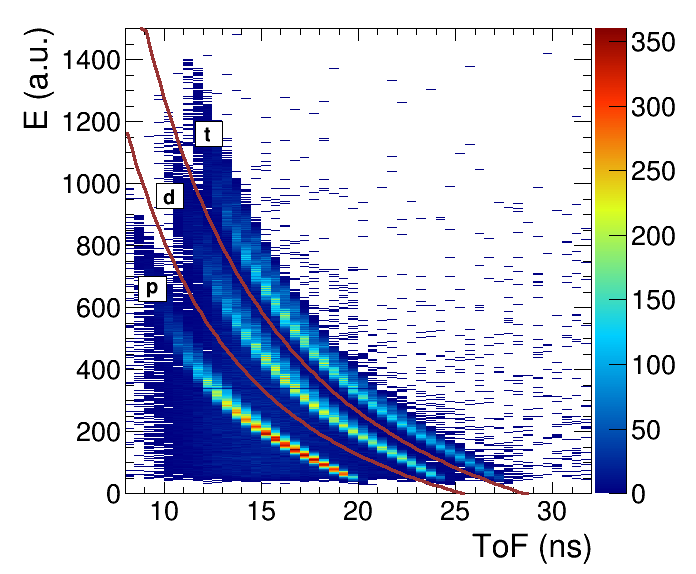}

\caption{\small{The correlations measured among the fragment discriminating variables are shown for He125 data at 10\degree. Top Left: the $\Delta E$ as a function of the $E$. Top Right: $E$ as a function of $\Delta t_{BGO-SC}$. Bottom Left: ratio between $\Delta E$ and $E$ as a function of the $ToF$. Bottom Right: $E$ as a function of the $ToF$.}}
\label{fig:PID}
\end{center}
\end{figure}

The correlations between the above variables are shown in Fig.~\ref{fig:PID} for the He125 data at 10\degree. In all the different combinations it is possible to clearly identify  three different event populations, related to the protons, deuterons and tritons. The most effective separation method is achieved when using $E$, the deposited energy in BGO, and $ToF$ as shown in Fig.~\ref{fig:PID}, Bottom Right plot. The red lines show the cuts used to separate the kinetic regions associated to the three different fragments.

\subsubsection{Energy measurement with BGO}

The energy calibration of BGO detectors has been performed by means of a dedicated data acquisition using proton and helium beams of different energies.
The light yield has been found to be proportional to the energy of the incoming particles in the kinetic energy ranges [48$\div$209 $\mega\electronvolt$] for protons and [50$\div$180 $\mega\electronvolt/u$] for $^4$He ions. The observed linearity, in the kinetic energy range of interest for the measurements presented in this paper, proved to be of great importance in order to obtain a good particle identification. However, since the fragment kinetic energy measurement derived from the $ToF$ is more accurate (see Sect. \ref{sec:tof}), the BGO calibrated energy was only used for particle ID purposes.

\subsubsection{Time of Flight}
\label{sec:tof}

The time needed by a fragment to travel the distance $L$ from STS$_1$ to STS$_2$ ($175\ \centi\meter$) is related to its kinetic energy $E_{kin}$ and its mass $m$ by the kinematic relation:

\begin{equation}
E_{kin}=m c^2 \left( \frac{1}{\sqrt{1- ( L / c \cdot ToF )^2}} - 1 \right)
\label{eq:ToF}
\end{equation}

Once the particle is identified and its mass is fixed, $E_{kin}$ can be evaluated from eq.~(\ref{eq:ToF}).


\section{Monte Carlo Simulation}
\label{sec:MC}

A fully detailed Monte Carlo (MC) simulation of the experimental setup has been performed by means of the FLUKA software (release 2011.2). The MC sample was hence used to evaluate the geometric and detection efficiency of the STS~(sec.~\ref{sec:stseff}) and the BGO~(sec. \ref{sec:bgoeff}) for the three different fragment populations and to tune the particle identification algorithms  (sec.~\ref{sec:mixing}).
The obtained results are fully independent of the nuclear models implemented within the FLUKA software as they rely only on the description of the interaction of the produced fragments with the PMMA and experimental setup matter. No assumption on the production spectra of the fragments was used to compute the efficiency. The only relevant physical processes were the fragment interactions with matter that are very well reproduced by the FLUKA software. 

\subsection{STS efficiency}
\label{sec:stseff}

The $ToF$ measurement requires a signal in both the STS detectors. Therefore the measured fragment yields has to be corrected to take into account the STS geometrical and detection efficiencies (the acceptance of the experimental setup will be discussed in the next section). The efficiency of each STS is derived as the ratio of the number of coincidences of the two STS and the number of events with a signal in the other STS in triggered events. The two scintillators are almost full efficient for charged particles but, given the its shape, in a sizable fraction of events a fragment hits laterally the BGO matrix without crossing the STS$_2$. From simple geometrical calculations this effect has been evaluated to give an inefficiency of about 20\%. This value has been then checked with a dedicated simulation. The disagreement between data and simulation is taken as systematic uncertainty. The measured product of STS$_1$ and STS$_2$ efficiencies (including the geometrical contribution) is $\varepsilon$=76$\pm$7\% for arm $a$ and $\varepsilon$=73$\pm$7\% for arms $b$ and $c$.

\subsection{BGO acceptance and efficiency}
\label{sec:bgoeff}

The BGO full detector efficiency, defined as the convolution of the geometrical acceptance of the experimental setup (related to the solid angle of the BGO matrix) and of the BGO detection efficiency, has been computed using a dedicated MC simulation. In order to take into account the production of the fragments than can occur along the beam path, protons, deuterons, and tritons with a simulated kinetic energy in the range of [50-250] $MeV/u$, are isotropically produced in a $12.65\ \centi\meter$ length empty cylindrical volume, aligned with the beam direction, with the radius corresponding to the beam size and ideally positioned within the PMMA. 

The product of the BGO acceptance and efficiency ($\varepsilon_{BGO}$ in the following) is computed as the fraction of $p$, $d$, and $t$ reaching the BGO matrix and releasing a signal above the trigger threshold. The statistical uncertainty on $\varepsilon_{BGO}$ is negligible. 

The systematic uncertainty is evaluated looking at the different fragment kinetic energies. The other geometry configurations ($10\ \centi\meter$ and $7.65\ \centi\meter$ target) can be obtained by imposing that the projectiles are emitted in the correct volume, however this different contribution ends up to a negligible effect.  The $\varepsilon_{BGO}$ dependence, mainly from geometrical acceptance, is on the angle of detection: $(4.9 \pm 0.2)\time 10^{-4}$ at $0\degree$ up to $(5.2 \pm 0.2)\time 10^{-4}$ at $30\degree$.

\subsection{Mixing of the fragment populations}	
\label{sec:mixing}

\begin{figure}[!ht]
\begin{center}
\includegraphics [width = 0.6 \textwidth] {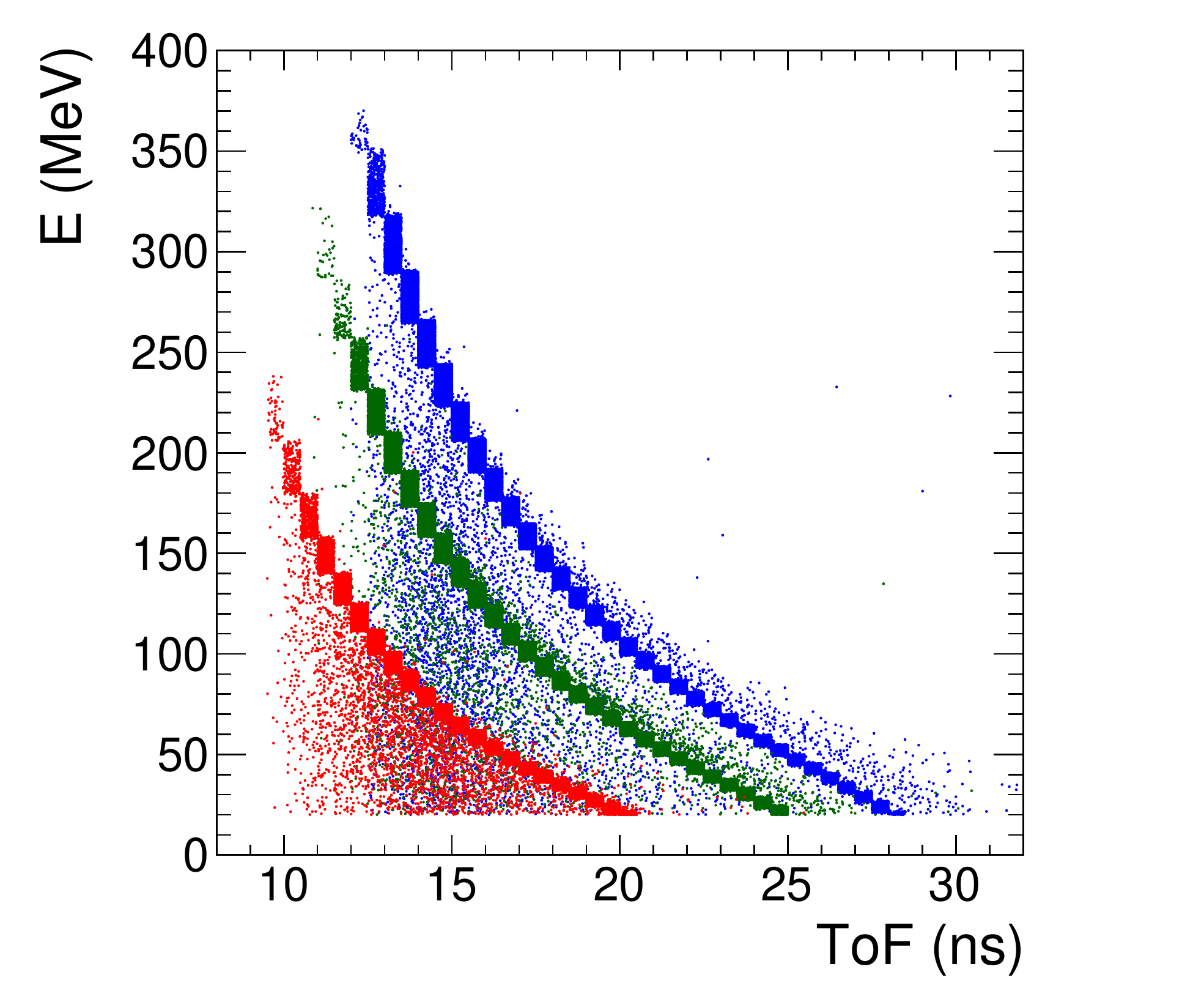}
\caption{\small{Example at $0\degree$ of the behavior of $E$ of a function of $ToF$ in MC events. The simulation is referred to the $125\ \mega\electronvolt/u$ $^4$He beam. The red, green and blue colors correspond respectively to protons, deuterons and tritons. The BGO energy cut at 20 MeV, to mimic the hardware threshold present in BGO read-out, is clearly visible.}}
\label{fig:bananaMC}
\end{center}
\end{figure}

The MC simulation has also been used to evaluate the efficiency of the particle identification algorithms used in the events selection (see Section~\ref{sec:pid}) and the contribution for possible cross feed of the different populations. As an example, Fig.~\ref{fig:bananaMC} shows the behavior of $E$ as a function of the $ToF$ for MC events in the $0\degree$ arm generated from the fragmentation of a $125\ \mega\electronvolt/u$~$^4$He beam. The $p$, $d$, and $t$ components are shown with different colors: in red protons, in green deuterons and in blue tritons. Three separate bands are clearly distinguishable.

The neutral component is not shown, being negligible. Indeed the ToF is computed using the STS and the probability of having photons or neutrons releasing hits in both the STS is very low; the MC predicts less than $1\permil$ background contamination. Therefore, the number of protons, deuterons, and tritons in data are hence obtained counting the number of particles in the regions defined as in Fig.~\ref{fig:PID} (Bottom Right).
 
In order to account for the mixing contribution, mainly due to particles the not fully contained in the BGO that release only part of their energy inside the BGO matrix, the probability for a fragment of type $i$ to be measured in the region $j$ of the $(E-ToF)$ plane (with $i,j = p,d,t$) was estimated as: $\varepsilon_{mix}^{ij} = N_{ij} / N_i$ that is the number of generated fragments of type $i$ measured in the region $j$ ($N_{ij}$) normalized to the total number of generated fragments $i$ in the $(E-ToF)$ plane ($N_i$). The values obtained for the mixing matrix ($\varepsilon_{mix}$) have been obtained using the MC simulation and are reported below:

\begin{center}                                                                                                                   
$\varepsilon_{mix} = \left( \begin{array}{ccc} \varepsilon^{pp} & \varepsilon^{pd} & \varepsilon^{pt}\\                                                                                                                                          
                                  \varepsilon^{dp} & \varepsilon^{dd} & \varepsilon^{dt}\\        
                                  \varepsilon^{tp} & \varepsilon^{td} & \varepsilon^{tt} \end{array} \right) = 
 \left( \begin{array}{rrr} 99.95\pm0.02 \% & 0.05\pm0.02 \% & --- \\                                                                                                                                          
                                  1.42\pm0.77 \% & 98.35\pm0.79 \% & 0.23\pm0.13 \% \\        
                                  0.99\pm0.59 \% & 1.32\pm0.34 \% & 97.69\pm0.88 \% \end{array} \right) $       
\end{center}                                                                                                                                                                                                      
The uncertainties are representing the spread of the matrix elements estimated from the simulation at different beam energies and different angles. As expected the largest contributions are coming from the misidentification of tritons in protons and deuterons or deuterons in protons that are related to the partial containment of the fragment inside the BGO detector, while the matrix elements above the diagonal (in which a protons is mistaken as a deuteron or a deuteron as a triton) are almost consistent with zero.

\section{Experimental Results}
\label{sec:exp}

The total fragment yield produced by the $^4$He beam on the PMMA target is given by the following formula:

\begin{equation} 
\frac{1}{N_{He}} \cdot \frac{dN_{frag}}{d\Omega} (sr^{-1}) = \frac{1}{4\pi} \cdot \frac{N_{frag}}{N_{He} \cdot c_{DS} \cdot (1-\delta_{DT}) \cdot \varepsilon_{STS} \cdot \varepsilon_{BGO}}
\label{eq:totyield}
\end{equation}

where:
\begin{itemize}
\item
$N_{frag}$ is the total number of detected fragments.
\item
$N_{He}$ is the number of beam particles impinging on the PMMA target corrected by the $C_{He}$ factor described at the end of Section~\ref{sec:rates}.
\item
$c_{DS}$ is the correction due to the trigger downscale at $0\degree$ and $5\degree$ mentioned in Section~\ref{sec:rates}, which is in the range [0.18-0.29].
\item
$\delta_{DT}$ is the inefficiency introduced by the DAQ dead time also described in Section~\ref{sec:rates} and evaluated comparing the VME scaler with the trigger rate acquired by the DAQ system. It ranges from from 20\% to 35\% depending on the beam rate.
\item
$\varepsilon_{STS}$ is the STS efficiency, including the geometrical inefficiency of STS$_2$ described in Section~\ref{sec:stseff}.
\item
$\varepsilon_{BGO}$ is the product of the BGO detection efficiency (including acceptance) as described in Section~\ref{sec:bgoeff}.
\end{itemize}

\begin{figure}[!ht]
\begin{center}
\includegraphics [width = 0.6 \textwidth] {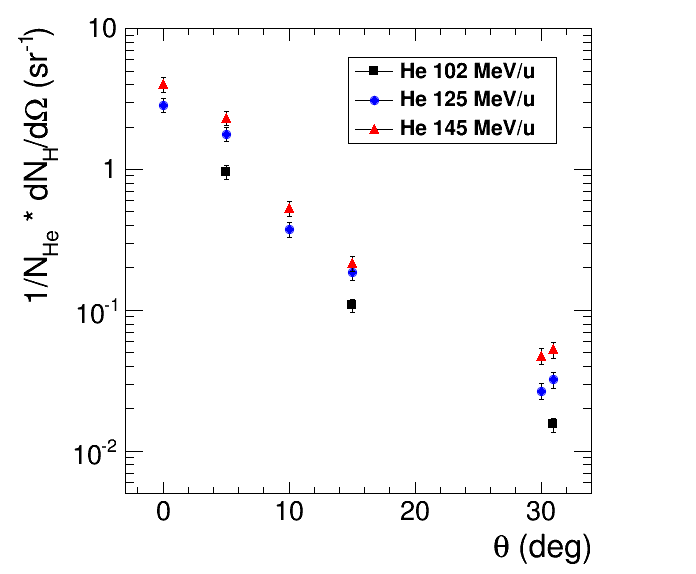}
\caption{\small{Total fragment yield ($p+d+t$), as a function of the detection angle for different beam energies. The $102\ \mega\electronvolt/u$ is not available for $0\degree$, $10\degree$ and $30\degree$ configurations (see text for details).}}
\label{fig:yieldVsAngle}
\end{center}
\end{figure}

Aim of this paper is the measurement of the flux of fragments escaping the PMMA: no matter correction is computed to take into account the low energy fragments absorbed in the PMMA.
The total fragment yield as a function of the detector angle is shown in Fig.~\ref{fig:yieldVsAngle} and reported in Table~\ref{tab:totalYield}. The uncertainty associated to the measured yield includes both the statistical and systematic contributions. Being the former well below 1\% it resulted to be negligible with respect to the latter. The main contribution to the systematic uncertainty comes from the STS efficiency (Section ~\ref{sec:stseff}): the corresponding uncertainty on the fragment yield is about 10\%. Other sources contributing to the systematic error are the uncertainty on the number of $^4$He beam primaries $N_{He}$ impinging on the PMMA target (4\%-6\% for the different beam configurations) and the error on $\varepsilon_{BGO}$ ($\sim$~4\%). Several additional contributions to the systematic error have been evaluated, as the error related to the uncertainty on the dead time efficiency and the uncertainty on $N_{frag}$ due to the BGO selection cut threshold, and proven to be totally negligible with respect to the main contributions described above. \\ 
For $0\degree$, $10\degree$ and $30\degree$ configurations the absolute yields are not shown for $102\ \mega\electronvolt/u$ energy data taking. This is due to the impossibility of normalizing the fragment yield: the number of incoming primary ions was not reliable in this data set, due to an hardware fault that affected the VME scaler counting. However this issue does not affect the relative yields, reported in section~\ref{sec:ryield}.
 
\begin{table}[ht]
\centering
\caption{Total fragment yield ($p+d+t$), measured by the STS and BGO detectors, for different beam energies at different angles. The uncertainties include both the statistical and the systematic contributions.}
\begin{tabular}{|c|c|c|c|c|c|c|}
\hline
beam energy & 0\degree & 5\degree & 10\degree & 15\degree & 30\degree & 30\degree$_{bis}$ \\
\cline{2-7}
($\mega\electronvolt/u$) & \multicolumn{2}{|c|}{(sr$^{-1}$)} & \multicolumn{2}{|c|}{(10$^{-1}$sr$^{-1}$)} & \multicolumn{2}{|c|}{(10$^{-2}$sr$^{-1}$)} \\
 \hline 
 $102$ & - & $1.0 \pm 0.1$ & - & $1.1 \pm 0.1 $ & - & $1.5 \pm 0.2$ \\ 
 \hline 
 $125$ & $2.9 \pm 0.3$ & $1.8 \pm 0.2$ & $3.8 \pm 0.4 $ & $1.8 \pm 0.2 $ & $2.7 \pm 0.3 $ & $3.2 \pm 0.4$ \\ 
 \hline 
 $145$ & $4.0 \pm 0.5$ & $2.3 \pm 0.3$ & $5.3 \pm 0.6 $ & $2.1 \pm 0.3 $ & $4.7 \pm 0.6 $ & $5.2 \pm 0.7$ \\
\hline
\end{tabular}
\label{tab:totalYield}
\end{table}

\subsection{Relative Yields}
\label{sec:ryield}

\begin{figure}[!ht]
\begin{center}
\includegraphics [width = 0.50 \textwidth] {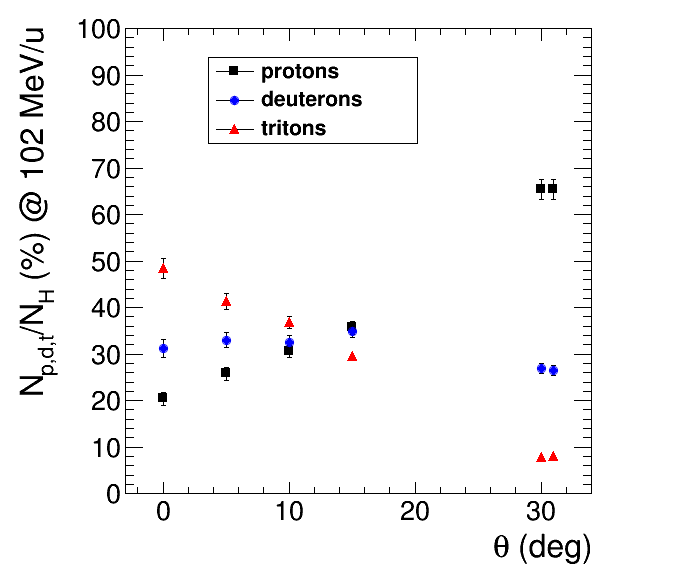}
\includegraphics [width = 0.50 \textwidth] {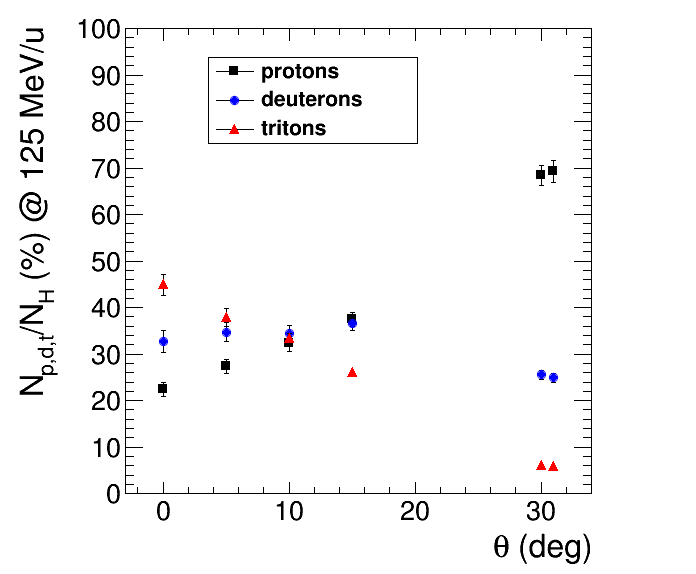}
\includegraphics [width = 0.50 \textwidth] {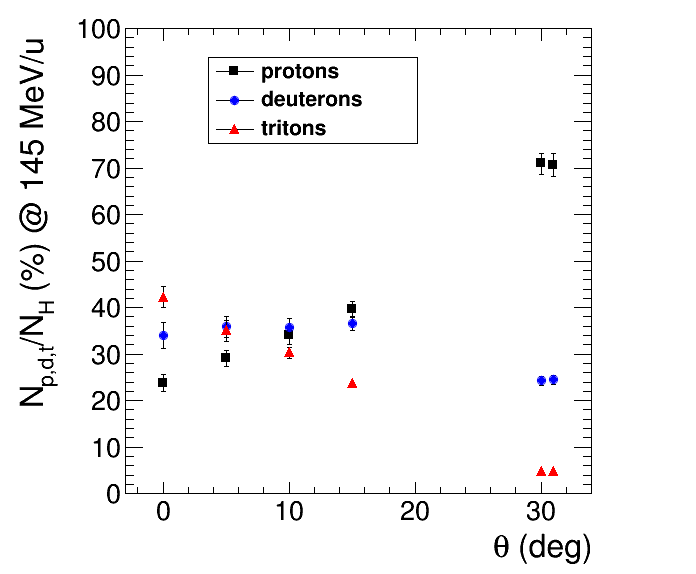}
\caption{\small{Fraction of protons, deuterons, and tritons normalized to the total number of H  fragments collected as a function of the detector angle for data collected using 102 MeV/u (top), 125 MeV/u (center) and 145 MeV/u (bottom) $^4$He beams.}}
\label{fig:relativeYieldVsAngle}
\end{center}
\end{figure}

The relative composition of protons, deuterons and tritons in the total fragment yield is measured by using the number of events counted separately in the three regions of the $E-ToF$ plane (defined as shown in Fig.~\ref{fig:PID}, Bottom Right), and normalized to the total number of fragments. The fraction of fragments shows a strong angular dependence, being the proton more abundant than deuterons, and tritons at large angle while the triton component is dominating at 0\degree. Conversely the dependence of the relative yields one the beam energy is quite small. A mixing correction is applied to take into account the small fraction of events placed in the wrong kinematic region due to a poor energy measurement in the BGO, as explained in Section~\ref{sec:mixing}. The relative $p$, $d$, and $t$ yields are reported in Table~\ref{tab:ryield} and shown in Fig.~\ref{fig:relativeYieldVsAngle} as a function of the angle and for different beam energies.
\begin{table}[ht]
\centering
\caption{Relative proton, deuteron and triton composition in the total yield (in \%), measured for different beam energies at different angles.}
\begin{tabular}{|c|c|c|c|c|c|c|}
\hline

$102\ \mega\electronvolt/u$ & 0\degree (\%) & 5\degree (\%) & 10\degree (\%) & 15\degree (\%) & 30\degree (\%) & 30\degree$_{bis}$ (\%) \\
\hline
\hline
proton  & 20.4 $\pm$ 1.4 & 25.8 $\pm$ 1.4 & 30.6 $\pm$ 1.3 & 35.6 $\pm$ 1.3 & 65.5 $\pm$ 2.2 & 65.4 $\pm$ 2.2 \\
deuteron  & 31.2 $\pm$ 1.9 & 33.0 $\pm$ 1.6 & 32.5 $\pm$ 1.4 & 34.9 $\pm$ 1.3 & 26.8 $\pm$ 1.1 & 26.5 $\pm$ 1.0 \\
triton  & 48.4 $\pm$ 2.2 & 41.3 $\pm$ 1.7 & 36.9 $\pm$ 1.3 & 29.4 $\pm$ 1.0 &  7.7 $\pm$ 0.4 &  8.0 $\pm$ 0.4 \\
\hline
$125\ \mega\electronvolt/u$ & 0\degree (\%) & 5\degree (\%) & 10\degree (\%) & 15\degree (\%) & 30\degree (\%) & 30\degree$_{bis}$ (\%) \\
\hline
\hline
proton  & 22.3 $\pm$ 1.5 & 27.4 $\pm$ 1.5 & 32.2 $\pm$ 1.5 & 37.5 $\pm$ 1.5 & 68.5 $\pm$ 2.2 & 69.3 $\pm$ 2.4 \\
deuteron  & 32.7 $\pm$ 2.3 & 34.8 $\pm$ 2.0 & 34.5 $\pm$ 1.6 & 36.5 $\pm$ 1.5 & 25.6 $\pm$ 1.0 & 24.9 $\pm$ 0.9 \\
triton  & 44.9 $\pm$ 2.2 & 37.9 $\pm$ 1.8 & 33.3 $\pm$ 1.2 & 26.0 $\pm$ 0.9 &  6.0 $\pm$ 0.3 &  5.8 $\pm$ 0.3 \\
\hline
$145\ \mega\electronvolt/u$ & 0\degree (\%) & 5\degree (\%) & 10\degree (\%) & 15\degree (\%) & 30\degree (\%) & 30\degree$_{bis}$ (\%) \\
\hline
\hline
proton  & 23.8 $\pm$ 1.8 & 29.0 $\pm$ 1.7 & 33.9 $\pm$ 1.8 & 39.7 $\pm$ 1.7 & 71.0 $\pm$ 2.3 & 70.7 $\pm$ 2.4 \\
deuteron  & 34.0 $\pm$ 2.7 & 35.9 $\pm$ 2.3 & 35.8 $\pm$ 1.8 & 36.6 $\pm$ 1.5 & 24.3 $\pm$ 1.0 & 24.5 $\pm$ 1.0 \\
triton  & 42.2 $\pm$ 2.3 & 35.0 $\pm$ 2.2 & 30.3 $\pm$ 1.2 & 23.8 $\pm$ 0.9 &  4.8 $\pm$ 0.3 &  4.8 $\pm$ 0.3 \\
\hline
\end{tabular}
\label{tab:ryield}
\end{table}

Some systematic contributions to the measurement of the total yield, as the uncertainties related to the STS efficiency and to the BGO acceptance, cancel out in the relative yield. However additional sources of systematic error arise in the separate evaluation of the number of the fragments:
\begin{itemize}
\item
 the uncertainty due to the partly arbitrary definition of the kinematic regions in the BGO-TOF plane (Fig.~\ref{fig:PID}, Bottom Right) is estimated, by moving the region borders among the proton/deuteron and deuteron/triton populations, to be in the range 3\%-6\% for all the $^4$He beam runs; 
\item
the error related to the uncertainty on the mixing matrix ($\varepsilon_{mix}$), obtained comparing in simulation samples the mixing size at different beam energies and angles, is less than 1\%.
\end{itemize}

The good agreement of the two results at 30\degree~(experimental configurations 30\degree~and 30\degree$_{bis}$) in Tables~\ref{tab:totalYield} and \ref{tab:ryield} expresses the reproducibility on the whole process of measurement.

\subsection{Energy spectra}
\label{sec:esp}

\begin{figure}[!ht]
\begin{center}
\includegraphics [width = 0.32 \textwidth] {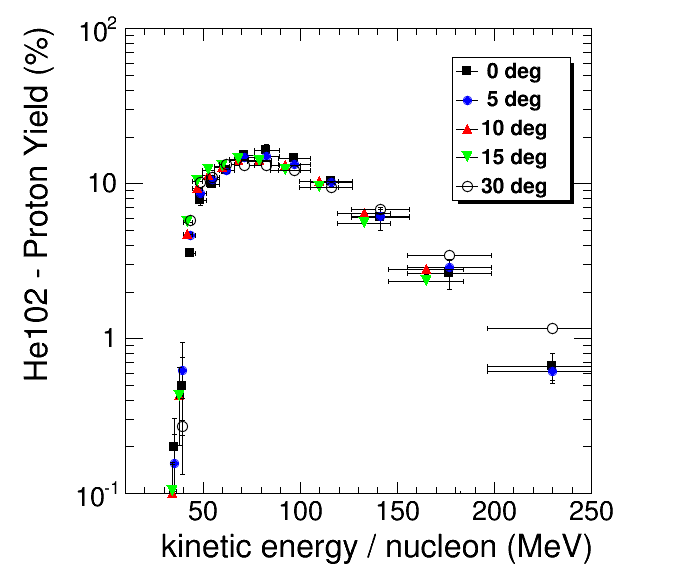}
\includegraphics [width = 0.32 \textwidth] {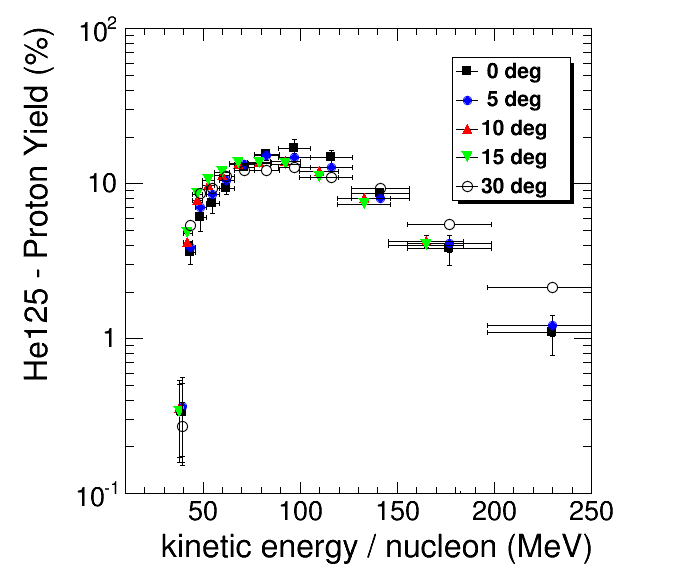}
\includegraphics [width = 0.32 \textwidth] {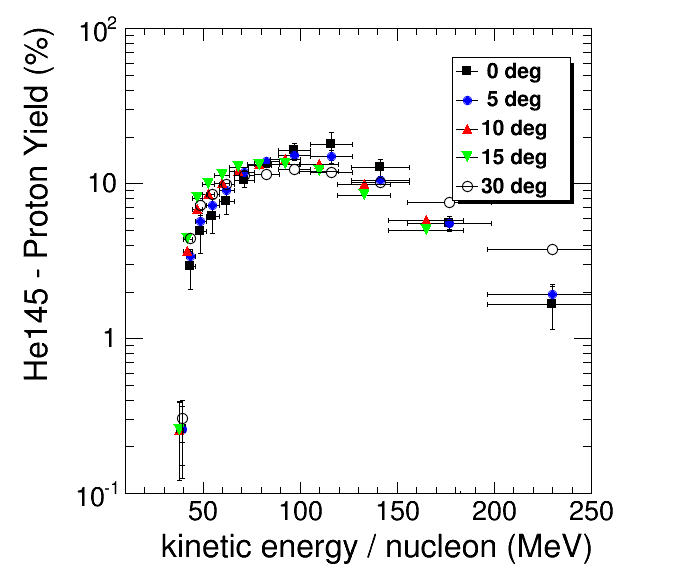}
\includegraphics [width = 0.32 \textwidth] {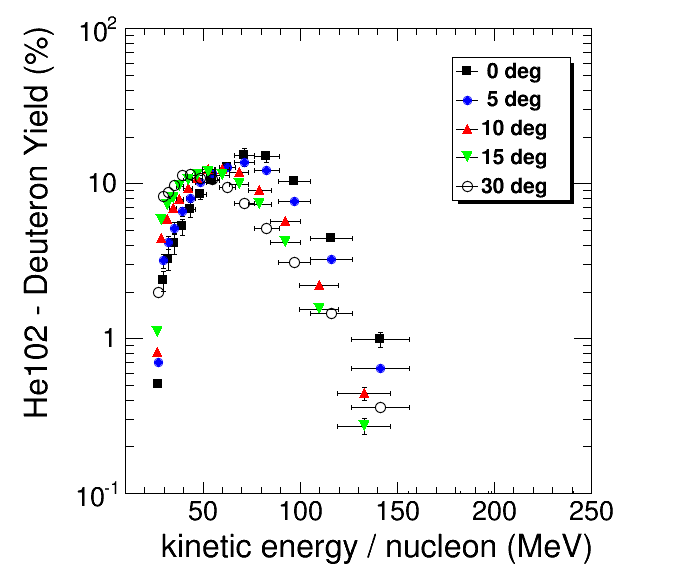}
\includegraphics [width = 0.32 \textwidth] {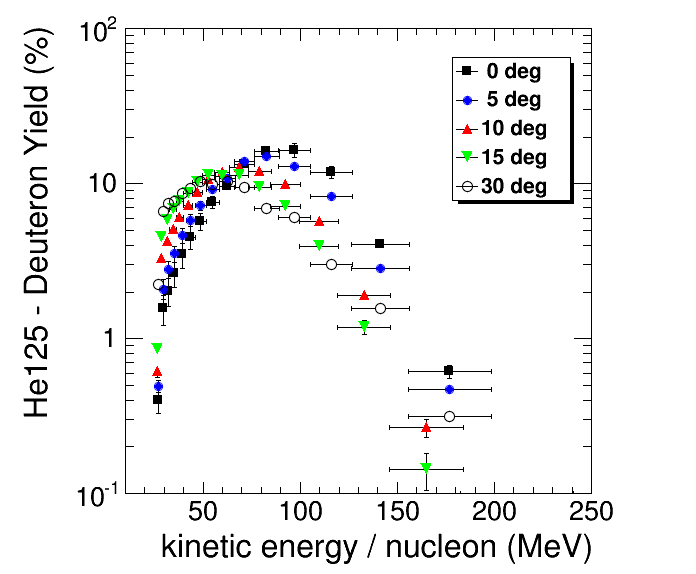}
\includegraphics [width = 0.32 \textwidth] {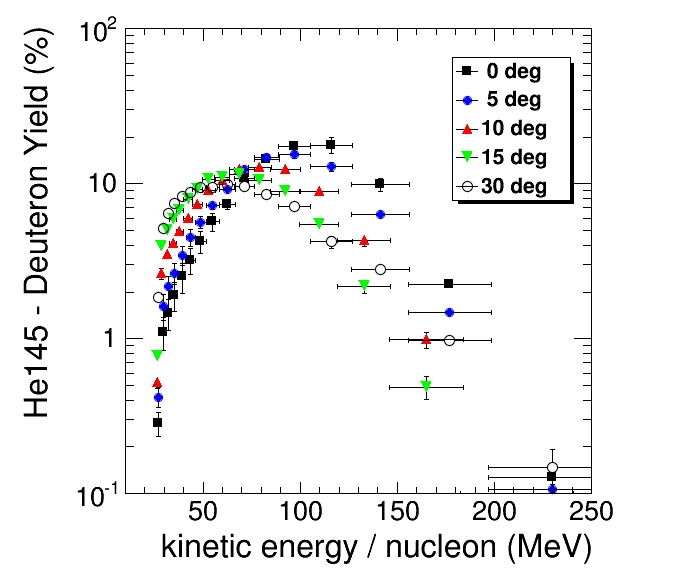}
\includegraphics [width = 0.32 \textwidth] {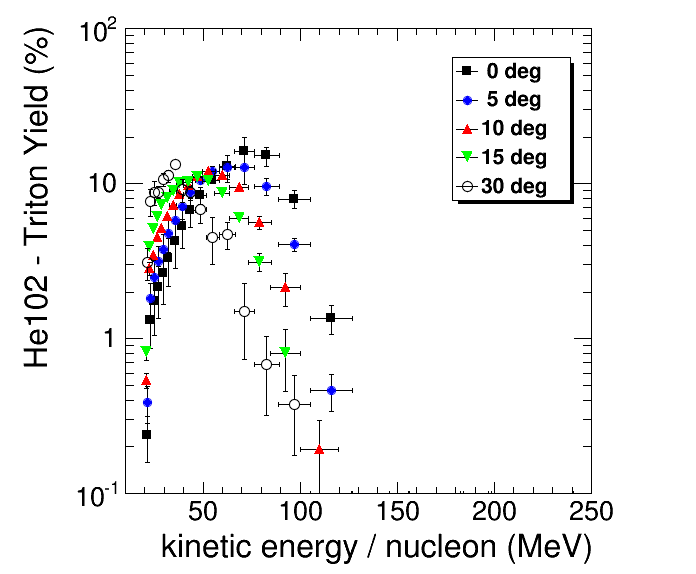}
\includegraphics [width = 0.32 \textwidth] {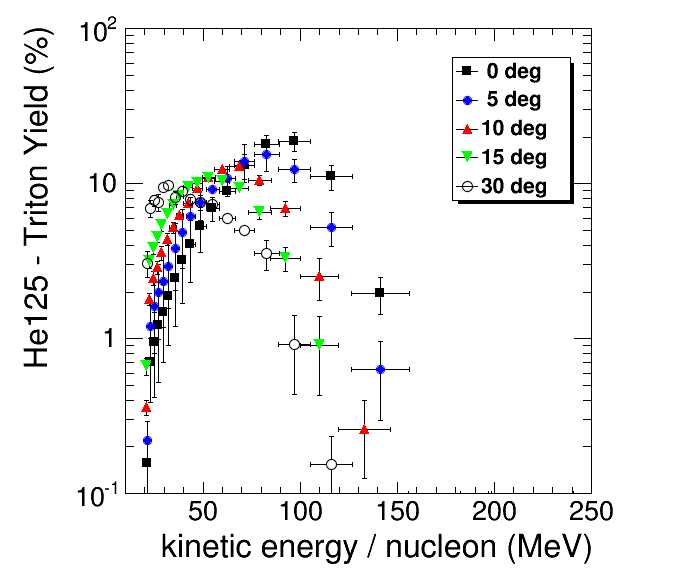}
\includegraphics [width = 0.32 \textwidth] {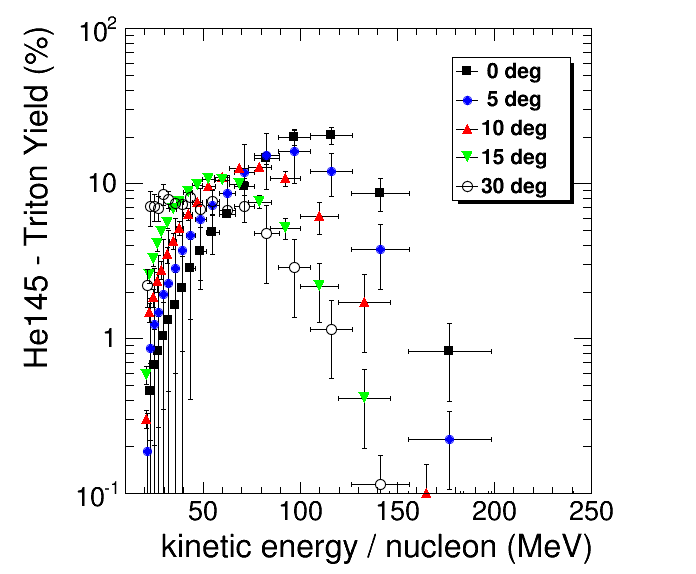}
\caption{\small{Relative yield of protons (top plots), deuterons (center), and tritons (bottom) as a function of nucleon kinetic energy for different angles with 102 MeV/u (left plots), 125 MeV/u (center) and 145 MeV/u (right) $^4$He beams.}}
\label{fig:YieldVsEnergy}
\end{center}
\end{figure}

The kinetic energy of the fragment was derived from the $ToF$ (eq.~\ref{eq:ToF}) being this method much more accurate than the measurement of the deposited energy in the BGO.
 The relative yields of the fragments as a function of kinetic energy per nucleon for different angles and beam energies are shown in Fig.~\ref{fig:YieldVsEnergy} and reported in the Tables~\ref{tab:yield_vs_energy_0deg}-\ref{tab:yield_vs_energy_30deg} in the following Appendix. The uncertainties on the kinetic energy per nucleon are obtained propagating a 0.5~ns ToF uncertainty through the eq.~(\ref{eq:ToF}). 

The differential fragment yield 
expressed in (MeV $\cdot$ sr)$^{-1}$ is given by:

\begin{equation} 
\frac{1}{N_{He}} \cdot \frac{d^2N_{p/d/t}}{dE d\Omega} = \left(\frac{1}{N_{He}} \cdot \frac{dN_{frag}}{d\Omega}\right) \cdot \left(\frac{N_{p/d/t}}{N_{frag}}\right) \cdot \left(\frac{1}{N_{p/d/t}} \cdot \frac{\Delta N_{p/d/t}(E)}{\Delta E}\right)
\label{eq:diffyield}
\end{equation}
where $1/N_{He} \cdot dN_{frag}/d\Omega$
 is the total yield given in Table~\ref{tab:totalYield}, 
 $N_{p/d/t}/N_{frag}$ 
 is the p/d/t fraction given in Table~\ref{tab:ryield}, and 
 $1/N_{p/d/t} \cdot \Delta N_{p/d/t}(E)/\Delta E$ 
 is the relative yield as a function of the nucleon kinetic energy listed in Tables~\ref{tab:yield_vs_energy_0deg}-\ref{tab:yield_vs_energy_30deg} and normalized to the energy bin size $\Delta E$ which is twice the uncertainty reported in the first column of the same tables.

\clearpage
\section{Conclusions}

This paper describes the results obtained with an experiment performed at the Heidelberg Ion-beam Therapy (HIT) center in Germany aiming at studying the fragmentation process of $^4$He interacting with a beam stopping PMMA target.\\
By means of a suitable set-up, the identity, the kinetic energy and the absolute flux of fragments produced in the interaction with the target were measured at five different angles: 0\degree, 5\degree, 10\degree, 15\degree, and 30\degree\;with respect to the beam direction. A dedicated Monte Carlo simulation was developed to evaluate in detail the detector efficiencies and their effect on the particle identification. Once calculated, these effects were used to correct the experimental results. \\
The detected fragments were, as expected, protons, deuterons and tritons. Three different beam energies were studied: $102\ \mega\electronvolt/u$, $125\ \mega\electronvolt/u$ and $145\ \mega\electronvolt/u$. For all energies the total fragment flux was found to rapidly decrease with the angle (an order of magnitude each $15\degree$). The relative yield of the different fragments shows two evident behaviors: 

\begin{itemize}
\item the lighter the fragment, the higher is its relative abundance at large angles; tritons are twice more abundant than protons at $0\degree$ while protons are about a factor 10 more abundant than tritons at $30\degree$. At $10\degree$ the yields of $p$, $d$ and $t$ contribute equally to the total flux. 
\item the heavy fragment component slightly decreases with the beam energy (e.g. at $10\degree$ triton component is 37.0\% at $102\ \mega\electronvolt/u$ and 30.6\% at $145\ \mega\electronvolt/u$).
\end{itemize}

The energy spectra of fragments were also studied. For all species an evident dependence of the spectrum maximum on the beam energy was found. For protons, spectra are quite large showing a tail above twice the beam energy with a small dependence on the angle. For deuterons and even more for tritons, spectra decrease rapidly for energies larger than the beam nominal one. This behavior is more accentuated at large angles.



\section*{Acknowledgments}
Authors would like to sincerely thank Marco Magi (Dipartimento di Scienze di Base e Applicate per l'Ingegneria, Sapienza Universit\`a di Roma) for his valuable effort in the construction of the several mechanical supports. This work has been partly supported by the Museo storico della Fisica e Centro di studi e ricerche Enrico Fermi. The access to the test beam at the Heidelberg Ion-beam Therapy center has been granted by the ULICE European program. We are indebted to Prof.~Dr.~Thomas Haberer and Dr.~Stephan Brons for having encouraged this measurement, made possible thanks to their support and to the help of the whole HIT staff.


\section*{References}

\bibliography{Frag_bib}

\appendix

\section{}
The relative yield of protons, deuterons, and tritons as a function of nucleon kinetic energy is reported for the different angles (Tables~\ref{tab:yield_vs_energy_0deg}-\ref{tab:yield_vs_energy_30deg}).
\begin{table}[ht]
\centering
\caption{Relative $p$, $d$, and $t$ yields as a function of nucleon kinetic energy at 0\degree}
\rotatebox{270}{
\small
\begin{tabular}{|c|r|r|r|r|r|r|r|r|r|}
\hline
\hline
 nucleon & \multicolumn{3}{c|}{He102 beam - 0\degree} & \multicolumn{3}{c|}{He125 beam - 0\degree} & \multicolumn{3}{c|}{He145 beam - 0\degree} \\
\cline{2-10}
 energy (MeV) & p(\%) & d (\%) & t(\%) & p(\%) & d (\%) & t(\%) & p(\%) & d (\%) & t(\%) \\ 
 \hline
 19.9 $\pm$ 0.7 & --- & --- & --- & --- & --- & --- & --- & --- & --- \\ 
 21.4 $\pm$ 0.8 & --- & --- & 0.2 $\pm$ 0.1 & --- & --- & 0.2 $\pm$ 0.1 & --- & --- & 0.1 $\pm$ 0.0 \\ 
 23.1 $\pm$ 0.9 & --- & --- & 1.3 $\pm$ 0.5 & --- & --- & 0.7 $\pm$ 0.3 & --- & --- & 0.5 $\pm$ 0.2 \\ 
 25.0 $\pm$ 1.0 & --- & --- & 1.7 $\pm$ 0.7 & --- & --- & 1.0 $\pm$ 0.5 & --- & --- & 0.7 $\pm$ 0.5 \\ 
 27.1 $\pm$ 1.1 & --- & 0.5 $\pm$ 0.0 & 2.1 $\pm$ 0.8 & --- & 0.4 $\pm$ 0.1 & 1.2 $\pm$ 0.7 & --- & 0.3 $\pm$ 0.0 & 0.8 $\pm$ 0.6 \\ 
 29.6 $\pm$ 1.3 & --- & 2.4 $\pm$ 0.3 & 2.6 $\pm$ 1.0 & --- & 1.6 $\pm$ 0.3 & 1.5 $\pm$ 0.8 & --- & 1.1 $\pm$ 0.3 & 1.0 $\pm$ 0.7 \\ 
 32.3 $\pm$ 1.5 & 0.1 $\pm$ 0.0 & 3.3 $\pm$ 0.5 & 3.3 $\pm$ 1.1 & --- & 2.0 $\pm$ 0.4 & 1.9 $\pm$ 1.0 & --- & 1.5 $\pm$ 0.3 & 1.3 $\pm$ 0.9 \\ 
 35.6 $\pm$ 1.7 & 0.2 $\pm$ 0.1 & 4.1 $\pm$ 0.6 & 4.2 $\pm$ 1.4 & 0.1 $\pm$ 0.0 & 2.6 $\pm$ 0.5 & 2.4 $\pm$ 1.2 & --- & 1.9 $\pm$ 0.4 & 1.6 $\pm$ 1.0 \\ 
 39.3 $\pm$ 2.0 & 0.5 $\pm$ 0.3 & 5.3 $\pm$ 0.6 & 5.3 $\pm$ 1.5 & 0.3 $\pm$ 0.2 & 3.5 $\pm$ 0.6 & 3.2 $\pm$ 1.5 & 0.3 $\pm$ 0.1 & 2.5 $\pm$ 0.5 & 2.1 $\pm$ 1.3 \\ 
 43.6 $\pm$ 2.4 & 3.5 $\pm$ 0.1 & 6.8 $\pm$ 0.7 & 6.7 $\pm$ 1.5 & 3.6 $\pm$ 0.6 & 4.5 $\pm$ 0.7 & 4.1 $\pm$ 1.7 & 2.9 $\pm$ 0.8 & 3.2 $\pm$ 0.6 & 2.8 $\pm$ 1.5 \\ 
 48.8 $\pm$ 2.8 & 7.8 $\pm$ 0.6 & 8.5 $\pm$ 0.5 & 8.4 $\pm$ 1.1 & 6.0 $\pm$ 1.1 & 5.7 $\pm$ 0.7 & 5.3 $\pm$ 1.7 & 4.9 $\pm$ 1.3 & 4.2 $\pm$ 0.7 & 3.6 $\pm$ 1.6 \\ 
 54.9 $\pm$ 3.4 & 9.9 $\pm$ 0.3 & 10.5 $\pm$ 0.1 & 10.5 $\pm$ 0.2 & 7.5 $\pm$ 1.0 & 7.6 $\pm$ 0.7 & 6.9 $\pm$ 1.2 & 6.1 $\pm$ 1.4 & 5.7 $\pm$ 0.7 & 4.9 $\pm$ 1.4 \\ 
 62.3 $\pm$ 4.1 & 12.2 $\pm$ 0.1 & 12.8 $\pm$ 0.7 & 12.9 $\pm$ 2.2 & 9.3 $\pm$ 0.7 & 9.6 $\pm$ 0.2 & 8.9 $\pm$ 0.6 & 7.7 $\pm$ 1.4 & 7.3 $\pm$ 0.5 & 6.4 $\pm$ 0.2 \\ 
 71.4 $\pm$ 5.1 & 15.3 $\pm$ 1.0 & 15.2 $\pm$ 1.5 & 16.2 $\pm$ 3.6 & 12.6 $\pm$ 0.2 & 13.3 $\pm$ 0.4 & 13.0 $\pm$ 2.4 & 10.4 $\pm$ 1.0 & 10.7 $\pm$ 0.2 & 9.7 $\pm$ 1.2 \\ 
 82.8 $\pm$ 6.4 & 16.4 $\pm$ 1.5 & 15.0 $\pm$ 1.2 & 15.1 $\pm$ 2.1 & 15.3 $\pm$ 1.1 & 16.2 $\pm$ 1.1 & 18.0 $\pm$ 2.5 & 13.4 $\pm$ 0.2 & 14.3 $\pm$ 0.2 & 14.6 $\pm$ 1.2 \\ 
 97.2 $\pm$ 8.2 & 14.5 $\pm$ 0.5 & 10.3 $\pm$ 0.3 & 8.0 $\pm$ 1.1 & 16.8 $\pm$ 2.4 & 16.4 $\pm$ 1.7 & 18.7 $\pm$ 2.7 & 16.3 $\pm$ 1.9 & 17.3 $\pm$ 1.3 & 19.9 $\pm$ 2.2 \\ 
 116.0 $\pm$ 10.9 & 10.3 $\pm$ 0.8 & 4.4 $\pm$ 0.2 & 1.4 $\pm$ 0.3 & 14.8 $\pm$ 1.6 & 11.8 $\pm$ 1.1 & 11.1 $\pm$ 2.0 & 17.8 $\pm$ 3.7 & 17.8 $\pm$ 2.0 & 20.6 $\pm$ 2.6 \\ 
 141.3 $\pm$ 14.9 & 6.0 $\pm$ 1.0 & 1.0 $\pm$ 0.1 & --- & 8.6 $\pm$ 0.5 & 4.1 $\pm$ 0.1 & 2.0 $\pm$ 0.5 & 12.7 $\pm$ 1.6 & 9.8 $\pm$ 0.9 & 8.7 $\pm$ 2.1 \\ 
 176.9 $\pm$ 21.5 & 2.6 $\pm$ 0.6 & 0.1 $\pm$ 0.0 & --- & 3.8 $\pm$ 0.8 & 0.6 $\pm$ 0.1 & --- & 5.5 $\pm$ 0.7 & 2.2 $\pm$ 0.0 & 0.8 $\pm$ 0.4 \\ 
 229.8 $\pm$ 33.2 & 0.7 $\pm$ 0.1 & --- & --- & 1.1 $\pm$ 0.3 & --- & --- & 1.7 $\pm$ 0.5 & 0.1 $\pm$ 0.0 & --- \\ 
 315.5 $\pm$ 57.3 & --- & --- & --- & 0.1 $\pm$ 0.0 & --- & --- & 0.2 $\pm$ 0.1 & --- & --- \\ 
\hline
\end{tabular}
}
\label{tab:yield_vs_energy_0deg}
\end{table}

\begin{table}[ht]
\centering
\caption{Relative $p$, $d$, and $t$ yields as a function of nucleon kinetic energy at 5\degree}
\rotatebox{270}{
\small
\begin{tabular}{|c|r|r|r|r|r|r|r|r|r|}
\hline
\hline
 nucleon & \multicolumn{3}{c|}{He102 beam - 5\degree} & \multicolumn{3}{c|}{He125 beam - 5\degree} & \multicolumn{3}{c|}{He145 beam - 5\degree} \\
\cline{2-10}
 energy (MeV) & p(\%) & d (\%) & t(\%) & p(\%) & d (\%) & t(\%) & p(\%) & d (\%) & t(\%) \\ 
\hline
 21.4 $\pm$ 0.8 & --- & --- & 0.4 $\pm$ 0.1 & --- & --- & 0.2 $\pm$ 0.1 & --- & --- & 0.2 $\pm$ 0.1 \\ 
 23.1 $\pm$ 0.9 & --- & --- & 1.8 $\pm$ 0.5 & --- & --- & 1.2 $\pm$ 0.5 & --- & --- & 0.9 $\pm$ 0.9 \\ 
 25.0 $\pm$ 1.0 & --- & --- & 2.5 $\pm$ 0.7 & --- & --- & 1.6 $\pm$ 0.8 & --- & --- & 1.2 $\pm$ 1.6 \\ 
 27.1 $\pm$ 1.1 & --- & 0.7 $\pm$ 0.0 & 3.1 $\pm$ 0.8 & --- & 0.5 $\pm$ 0.0 & 2.0 $\pm$ 1.0 & --- & 0.4 $\pm$ 0.1 & 1.5 $\pm$ 1.9 \\ 
 29.6 $\pm$ 1.3 & --- & 3.2 $\pm$ 0.3 & 3.8 $\pm$ 0.9 & --- & 2.1 $\pm$ 0.3 & 2.3 $\pm$ 1.1 & --- & 1.6 $\pm$ 0.3 & 1.9 $\pm$ 2.4 \\ 
 32.3 $\pm$ 1.5 & --- & 4.2 $\pm$ 0.4 & 4.8 $\pm$ 1.1 & --- & 2.8 $\pm$ 0.3 & 2.9 $\pm$ 1.4 & --- & 2.2 $\pm$ 0.4 & 2.3 $\pm$ 2.7 \\ 
 35.6 $\pm$ 1.7 & 0.2 $\pm$ 0.1 & 5.2 $\pm$ 0.4 & 5.8 $\pm$ 1.2 & 0.1 $\pm$ 0.0 & 3.5 $\pm$ 0.4 & 3.8 $\pm$ 1.7 & 0.1 $\pm$ 0.0 & 2.6 $\pm$ 0.4 & 2.8 $\pm$ 3.2 \\ 
 39.3 $\pm$ 2.0 & 0.6 $\pm$ 0.3 & 6.6 $\pm$ 0.4 & 7.1 $\pm$ 1.3 & 0.4 $\pm$ 0.2 & 4.6 $\pm$ 0.5 & 4.8 $\pm$ 2.0 & 0.3 $\pm$ 0.1 & 3.5 $\pm$ 0.5 & 3.7 $\pm$ 3.9 \\ 
 43.6 $\pm$ 2.4 & 4.7 $\pm$ 0.2 & 8.1 $\pm$ 0.2 & 8.7 $\pm$ 1.0 & 3.9 $\pm$ 0.2 & 5.8 $\pm$ 0.5 & 6.1 $\pm$ 2.1 & 3.4 $\pm$ 0.4 & 4.5 $\pm$ 0.6 & 4.6 $\pm$ 4.2 \\ 
 48.8 $\pm$ 2.8 & 8.6 $\pm$ 0.1 & 10.1 $\pm$ 0.2 & 10.4 $\pm$ 0.3 & 7.0 $\pm$ 0.5 & 7.2 $\pm$ 0.5 & 7.5 $\pm$ 1.7 & 5.7 $\pm$ 0.7 & 5.6 $\pm$ 0.5 & 5.8 $\pm$ 3.5 \\ 
 54.9 $\pm$ 3.4 & 10.6 $\pm$ 0.0 & 11.7 $\pm$ 0.2 & 12.1 $\pm$ 0.9 & 8.5 $\pm$ 0.4 & 9.1 $\pm$ 0.2 & 9.2 $\pm$ 0.4 & 7.3 $\pm$ 0.7 & 7.2 $\pm$ 0.4 & 7.2 $\pm$ 2.0 \\ 
 62.3 $\pm$ 4.1 & 12.1 $\pm$ 0.3 & 12.8 $\pm$ 0.7 & 12.7 $\pm$ 2.5 & 10.5 $\pm$ 0.1 & 10.8 $\pm$ 0.3 & 10.8 $\pm$ 2.0 & 9.0 $\pm$ 0.5 & 9.2 $\pm$ 0.0 & 8.7 $\pm$ 2.5 \\ 
 71.4 $\pm$ 5.1 & 14.8 $\pm$ 0.5 & 13.7 $\pm$ 0.8 & 12.8 $\pm$ 2.9 & 13.4 $\pm$ 0.3 & 14.0 $\pm$ 0.7 & 14.0 $\pm$ 3.8 & 11.8 $\pm$ 0.2 & 12.3 $\pm$ 0.4 & 11.8 $\pm$ 6.2 \\ 
 82.8 $\pm$ 6.4 & 15.1 $\pm$ 0.5 & 12.1 $\pm$ 0.4 & 9.6 $\pm$ 1.2 & 15.1 $\pm$ 0.7 & 15.0 $\pm$ 0.9 & 15.3 $\pm$ 3.4 & 14.0 $\pm$ 0.4 & 14.6 $\pm$ 0.7 & 15.2 $\pm$ 6.0 \\ 
 97.2 $\pm$ 8.2 & 13.3 $\pm$ 0.1 & 7.7 $\pm$ 0.0 & 4.0 $\pm$ 0.4 & 14.8 $\pm$ 0.9 & 13.0 $\pm$ 0.7 & 12.3 $\pm$ 2.1 & 15.3 $\pm$ 1.2 & 15.5 $\pm$ 1.0 & 16.1 $\pm$ 6.2 \\ 
 116.0 $\pm$ 10.9 & 10.2 $\pm$ 0.6 & 3.3 $\pm$ 0.1 & 0.5 $\pm$ 0.1 & 12.7 $\pm$ 0.3 & 8.3 $\pm$ 0.2 & 5.2 $\pm$ 1.3 & 14.9 $\pm$ 1.4 & 12.8 $\pm$ 0.8 & 12.0 $\pm$ 3.7 \\ 
 141.3 $\pm$ 14.9 & 6.2 $\pm$ 0.7 & 0.6 $\pm$ 0.0 & --- & 8.1 $\pm$ 0.4 & 2.8 $\pm$ 0.0 & 0.6 $\pm$ 0.3 & 10.5 $\pm$ 0.2 & 6.3 $\pm$ 0.2 & 3.8 $\pm$ 1.7 \\ 
 176.9 $\pm$ 21.5 & 2.9 $\pm$ 0.4 & --- & --- & 4.1 $\pm$ 0.5 & 0.5 $\pm$ 0.0 & --- & 5.6 $\pm$ 0.5 & 1.5 $\pm$ 0.0 & 0.2 $\pm$ 0.1 \\ 
 229.8 $\pm$ 33.2 & 0.6 $\pm$ 0.1 & --- & --- & 1.2 $\pm$ 0.2 & --- & --- & 1.9 $\pm$ 0.3 & 0.1 $\pm$ 0.0 & --- \\ 
 315.5 $\pm$ 57.3 & 0.1 $\pm$ 0.0 & --- & --- & 0.2 $\pm$ 0.0 & --- & --- & 0.3 $\pm$ 0.0 & --- & --- \\ 
\hline
\end{tabular}
}
\label{tab:yield_vs_energy_5deg}
\end{table}

\begin{table}[ht]
\centering
\caption{Relative $p$, $d$, and $t$ yields as a function of nucleon kinetic energy at 10\degree}
\rotatebox{270}{
\small
\begin{tabular}{|c|r|r|r|r|r|r|r|r|r|}
\hline
\hline
 nucleon & \multicolumn{3}{c|}{He102 beam - 10\degree} & \multicolumn{3}{c|}{He125 beam - 10\degree} & \multicolumn{3}{c|}{He145 beam - 10\degree} \\
\cline{2-10}
 energy (MeV) & p(\%) & d (\%) & t(\%) & p(\%) & d (\%) & t(\%) & p(\%) & d (\%) & t(\%) \\ 
\hline
 20.9 $\pm$ 0.8 & --- & --- & 0.5 $\pm$ 0.1 & --- & --- & 0.4 $\pm$ 0.0 & --- & --- & 0.3 $\pm$ 0.0 \\ 
 22.6 $\pm$ 0.9 & --- & --- & 2.8 $\pm$ 0.3 & --- & --- & 1.8 $\pm$ 0.2 & --- & --- & 1.5 $\pm$ 0.2 \\ 
 24.4 $\pm$ 1.0 & --- & --- & 3.5 $\pm$ 0.3 & --- & --- & 2.4 $\pm$ 0.3 & --- & --- & 1.8 $\pm$ 0.3 \\ 
 26.5 $\pm$ 1.1 & --- & 0.8 $\pm$ 0.1 & 4.5 $\pm$ 0.3 & --- & 0.6 $\pm$ 0.0 & 2.9 $\pm$ 0.3 & --- & 0.5 $\pm$ 0.0 & 2.3 $\pm$ 0.3 \\ 
 28.8 $\pm$ 1.3 & --- & 4.5 $\pm$ 0.3 & 5.1 $\pm$ 0.3 & --- & 3.3 $\pm$ 0.2 & 3.6 $\pm$ 0.4 & --- & 2.6 $\pm$ 0.2 & 2.8 $\pm$ 0.4 \\ 
 31.5 $\pm$ 1.4 & --- & 5.9 $\pm$ 0.3 & 6.2 $\pm$ 0.3 & --- & 4.3 $\pm$ 0.3 & 4.4 $\pm$ 0.4 & --- & 3.5 $\pm$ 0.2 & 3.5 $\pm$ 0.4 \\ 
 34.5 $\pm$ 1.7 & 0.1 $\pm$ 0.1 & 6.9 $\pm$ 0.2 & 7.3 $\pm$ 0.3 & 0.1 $\pm$ 0.0 & 5.1 $\pm$ 0.2 & 5.2 $\pm$ 0.4 & --- & 4.1 $\pm$ 0.3 & 4.3 $\pm$ 0.5 \\ 
 38.1 $\pm$ 1.9 & 0.4 $\pm$ 0.2 & 8.0 $\pm$ 0.2 & 8.5 $\pm$ 0.2 & 0.4 $\pm$ 0.2 & 6.0 $\pm$ 0.2 & 6.2 $\pm$ 0.4 & 0.3 $\pm$ 0.1 & 4.9 $\pm$ 0.3 & 5.1 $\pm$ 0.5 \\ 
 42.3 $\pm$ 2.2 & 4.7 $\pm$ 0.2 & 9.3 $\pm$ 0.1 & 9.7 $\pm$ 0.2 & 4.2 $\pm$ 0.1 & 7.2 $\pm$ 0.2 & 7.5 $\pm$ 0.4 & 3.7 $\pm$ 0.2 & 6.0 $\pm$ 0.3 & 6.3 $\pm$ 0.6 \\ 
 47.1 $\pm$ 2.7 & 9.3 $\pm$ 0.0 & 10.8 $\pm$ 0.0 & 11.2 $\pm$ 0.1 & 7.8 $\pm$ 0.2 & 8.8 $\pm$ 0.2 & 9.3 $\pm$ 0.3 & 6.8 $\pm$ 0.3 & 7.3 $\pm$ 0.3 & 7.6 $\pm$ 0.5 \\ 
 53.0 $\pm$ 3.2 & 11.3 $\pm$ 0.2 & 12.4 $\pm$ 0.1 & 12.1 $\pm$ 0.2 & 9.8 $\pm$ 0.2 & 10.6 $\pm$ 0.1 & 11.0 $\pm$ 0.1 & 8.6 $\pm$ 0.3 & 9.0 $\pm$ 0.2 & 9.6 $\pm$ 0.5 \\ 
 60.0 $\pm$ 3.9 & 12.8 $\pm$ 0.3 & 12.3 $\pm$ 0.2 & 11.3 $\pm$ 0.4 & 11.3 $\pm$ 0.2 & 11.7 $\pm$ 0.1 & 12.3 $\pm$ 0.2 & 10.0 $\pm$ 0.1 & 10.5 $\pm$ 0.1 & 11.0 $\pm$ 0.2 \\ 
 68.5 $\pm$ 4.7 & 14.2 $\pm$ 0.3 & 11.8 $\pm$ 0.2 & 9.5 $\pm$ 0.4 & 13.3 $\pm$ 0.2 & 12.7 $\pm$ 0.2 & 12.9 $\pm$ 0.3 & 12.0 $\pm$ 0.1 & 12.3 $\pm$ 0.1 & 12.5 $\pm$ 0.2 \\ 
 79.1 $\pm$ 5.9 & 14.1 $\pm$ 0.1 & 9.0 $\pm$ 0.2 & 5.6 $\pm$ 0.5 & 13.6 $\pm$ 0.4 & 12.0 $\pm$ 0.4 & 10.5 $\pm$ 0.8 & 13.2 $\pm$ 0.3 & 12.7 $\pm$ 0.3 & 12.7 $\pm$ 0.6 \\ 
 92.5 $\pm$ 7.6 & 13.0 $\pm$ 0.2 & 5.7 $\pm$ 0.2 & 2.1 $\pm$ 0.5 & 14.0 $\pm$ 0.3 & 9.9 $\pm$ 0.3 & 6.9 $\pm$ 0.8 & 14.2 $\pm$ 0.6 & 12.3 $\pm$ 0.5 & 10.7 $\pm$ 1.2 \\ 
 109.8 $\pm$ 10.0 & 10.3 $\pm$ 0.5 & 2.2 $\pm$ 0.1 & 0.2 $\pm$ 0.1 & 12.0 $\pm$ 0.1 & 5.7 $\pm$ 0.3 & 2.5 $\pm$ 0.8 & 13.3 $\pm$ 0.4 & 9.0 $\pm$ 0.5 & 6.2 $\pm$ 1.4 \\ 
 132.9 $\pm$ 13.5 & 6.4 $\pm$ 0.4 & 0.4 $\pm$ 0.0 & --- & 8.0 $\pm$ 0.4 & 1.9 $\pm$ 0.1 & 0.3 $\pm$ 0.1 & 9.8 $\pm$ 0.1 & 4.3 $\pm$ 0.3 & 1.7 $\pm$ 0.9 \\ 
 164.8 $\pm$ 19.2 & 2.8 $\pm$ 0.2 & --- & --- & 4.3 $\pm$ 0.4 & 0.3 $\pm$ 0.0 & --- & 5.8 $\pm$ 0.4 & 1.0 $\pm$ 0.1 & 0.1 $\pm$ 0.1 \\ 
\hline
\end{tabular}
}
\label{tab:yield_vs_energy_10deg}
\end{table}

\begin{table}[ht]
\centering
\caption{Relative $p$, $d$, and $t$ yields as a function of nucleon kinetic energy at 15\degree}
\rotatebox{270}{
\small
\begin{tabular}{|c|r|r|r|r|r|r|r|r|r|}
\hline
\hline
 nucleon & \multicolumn{3}{c|}{He102 beam - 15\degree} & \multicolumn{3}{c|}{He125 beam - 15\degree} & \multicolumn{3}{c|}{He145 beam - 15\degree} \\
\cline{2-10}
  energy (MeV) & p(\%) & d (\%) & t(\%) & p(\%) & d (\%) & t(\%) & p(\%) & d (\%) & t(\%) \\ 
\hline
 20.9 $\pm$ 0.8 &  --- &  --- &  0.8 $\pm$ 0.1 &  --- &  --- &  0.7 $\pm$ 0.1 &  --- &  --- &  0.6 $\pm$ 0.1 \\ 
 22.6 $\pm$ 0.9 &  --- &  --- &  3.9 $\pm$ 0.2 &  --- &  --- &  3.1 $\pm$ 0.3 &  --- &  --- &  2.5 $\pm$ 0.3 \\ 
 24.4 $\pm$ 1.0 &  --- &  --- &  5.1 $\pm$ 0.3 &  --- &  --- &  3.8 $\pm$ 0.3 &  --- &  --- &  3.2 $\pm$ 0.3 \\ 
 26.5 $\pm$ 1.1 &  --- &  1.1 $\pm$ 0.1 &  6.1 $\pm$ 0.3 &  --- &  0.8 $\pm$ 0.0 &  4.5 $\pm$ 0.3 &  --- &  0.8 $\pm$ 0.0 &  4.0 $\pm$ 0.4 \\ 
 28.8 $\pm$ 1.3 &  --- & 5.8 $\pm$ 0.2 & 7.2 $\pm$ 0.4 & --- & 4.5 $\pm$ 0.2 & 5.3 $\pm$ 0.4 & --- & 4.0 $\pm$ 0.2 & 4.8 $\pm$ 0.4 \\ 
 31.5 $\pm$ 1.4 & --- & 7.3 $\pm$ 0.2 & 8.0 $\pm$ 0.3 & --- & 5.8 $\pm$ 0.2 & 6.4 $\pm$ 0.4 & --- & 5.0 $\pm$ 0.2 & 5.6 $\pm$ 0.4 \\ 
 34.5 $\pm$ 1.7 & 0.1 $\pm$ 0.1 & 8.1 $\pm$ 0.1 & 8.9 $\pm$ 0.3 & 0.1 $\pm$ 0.0 & 6.8 $\pm$ 0.2 & 7.2 $\pm$ 0.4 & 0.1 $\pm$ 0.0 & 5.9 $\pm$ 0.2 & 6.8 $\pm$ 0.5 \\ 
 38.1 $\pm$ 1.9 & 0.4 $\pm$ 0.2 & 9.6 $\pm$ 0.2 & 10.1 $\pm$ 0.2 & 0.3 $\pm$ 0.2 & 7.6 $\pm$ 0.1 & 8.2 $\pm$ 0.3 & 0.3 $\pm$ 0.1 & 6.7 $\pm$ 0.2 & 7.6 $\pm$ 0.4 \\ 
 42.3 $\pm$ 2.2 & 5.6 $\pm$ 0.2 & 10.5 $\pm$ 0.1 & 10.2 $\pm$ 0.1 & 4.8 $\pm$ 0.1 & 8.6 $\pm$ 0.1 & 9.4 $\pm$ 0.2 & 4.4 $\pm$ 0.1 & 8.0 $\pm$ 0.1 & 8.8 $\pm$ 0.3 \\ 
 47.1 $\pm$ 2.7 & 10.4 $\pm$ 0.0 & 11.3 $\pm$ 0.0 & 11.0 $\pm$ 0.2 & 8.6 $\pm$ 0.0 & 10.1 $\pm$ 0.1 & 1--- & 8.1 $\pm$ 0.1 & 9.2 $\pm$ 0.1 & 9.7 $\pm$ 0.2 \\ 
 53.0 $\pm$ 3.2 & 12.2 $\pm$ 0.1 & 11.8 $\pm$ 0.1 & 10.3 $\pm$ 0.3 & 10.5 $\pm$ 0.0 & 11.3 $\pm$ 0.0 & 10.8 $\pm$ 0.2 & 9.8 $\pm$ 0.1 & 10.7 $\pm$ 0.1 & 10.6 $\pm$ 0.0 \\ 
 60.0 $\pm$ 3.9 & 12.9 $\pm$ 0.2 & 11.3 $\pm$ 0.1 & 8.6 $\pm$ 0.3 & 11.9 $\pm$ 0.1 & 11.1 $\pm$ 0.1 & 10.4 $\pm$ 0.3 & 11.4 $\pm$ 0.1 & 10.9 $\pm$ 0.1 & 10.5 $\pm$ 0.2 \\ 
 68.5 $\pm$ 4.7 & 14.4 $\pm$ 0.1 & 9.9 $\pm$ 0.1 & 5.9 $\pm$ 0.4 & 13.5 $\pm$ 0.2 & 11.4 $\pm$ 0.2 & 9.3 $\pm$ 0.5 & 12.7 $\pm$ 0.1 & 11.4 $\pm$ 0.1 & 9.9 $\pm$ 0.4 \\ 
 79.1 $\pm$ 5.9 & 13.9 $\pm$ 0.1 & 7.4 $\pm$ 0.1 & 3.1 $\pm$ 0.4 & 13.5 $\pm$ 0.1 & 9.4 $\pm$ 0.2 & 6.5 $\pm$ 0.6 & 13.1 $\pm$ 0.1 & 10.5 $\pm$ 0.2 & 7.6 $\pm$ 0.7 \\ 
 92.5 $\pm$ 7.6 & 12.2 $\pm$ 0.1 & 4.2 $\pm$ 0.1 & 0.8 $\pm$ 0.3 & 13.4 $\pm$ 0.0 & 7.2 $\pm$ 0.2 & 3.3 $\pm$ 0.6 & 13.2 $\pm$ 0.2 & 8.8 $\pm$ 0.2 & 5.2 $\pm$ 0.8 \\ 
 109.8 $\pm$ 10.0 & 9.5 $\pm$ 0.3 & 1.6 $\pm$ 0.1 & --- & 11.0 $\pm$ 0.1 & 3.9 $\pm$ 0.2 & 0.9 $\pm$ 0.5 & 12.0 $\pm$ 0.0 & 5.4 $\pm$ 0.3 & 2.2 $\pm$ 0.9 \\ 
 132.9 $\pm$ 13.5 & 5.5 $\pm$ 0.2 & 0.3 $\pm$ 0.0 & --- & 7.4 $\pm$ 0.2 & 1.2 $\pm$ 0.1 & 0.1 $\pm$ 0.0 & 8.3 $\pm$ 0.1 & 2.2 $\pm$ 0.2 & 0.4 $\pm$ 0.2 \\ 
 164.8 $\pm$ 19.2 & 2.3 $\pm$ 0.1 & --- & --- & 4.0 $\pm$ 0.2 & 0.1 $\pm$ 0.0 & --- & 5.0 $\pm$ 0.2 & 0.5 $\pm$ 0.1 & --- \\ 
\hline
\end{tabular}
}
\label{tab:yield_vs_energy_15deg}
\end{table}

\begin{table}[ht]
\centering
\caption{Relative $p$, $d$, and $t$ yields as a function of nucleon kinetic energy at 30\degree}
\rotatebox{270}{
\small
\begin{tabular}{|c|r|r|r|r|r|r|r|r|r|}
\hline
\hline
 nucleon & \multicolumn{3}{c|}{He102 beam - 30\degree} & \multicolumn{3}{c|}{He125 beam - 30\degree} & \multicolumn{3}{c|}{He145 beam - 30\degree} \\
\cline{2-10}
 energy (MeV) & p(\%) & d (\%) & t(\%) & p(\%) & d (\%) & t(\%) & p(\%) & d (\%) & t(\%) \\ 
\hline
 21.4 $\pm$ 0.8 & --- & --- & 3.1 $\pm$ 0.7 & --- & --- & 3.1 $\pm$ 0.6 & --- & --- & 2.2 $\pm$ 0.6 \\ 
 23.1 $\pm$ 0.9 & --- & --- & 7.7 $\pm$ 1.5 & --- & --- & 6.9 $\pm$ 0.8 & --- & --- & 7.1 $\pm$ 1.8 \\ 
 25.0 $\pm$ 1.0 & --- & --- & 8.7 $\pm$ 1.5 & --- & --- & 7.8 $\pm$ 0.7 & --- & --- & 7.1 $\pm$ 1.4 \\ 
 27.1 $\pm$ 1.1 & --- & 2.0 $\pm$ 0.1 & 8.8 $\pm$ 0.8 & --- & 2.2 $\pm$ 0.1 & 7.6 $\pm$ 1.0 & --- & 1.9 $\pm$ 0.1 & 7.0 $\pm$ 1.3 \\ 
 29.6 $\pm$ 1.3 & --- & 8.3 $\pm$ 0.4 & 10.6 $\pm$ 1.0 & --- & 6.6 $\pm$ 0.2 & 9.5 $\pm$ 0.5 & --- & 5.2 $\pm$ 0.3 & 8.5 $\pm$ 1.3 \\ 
 32.3 $\pm$ 1.5 & --- & 8.8 $\pm$ 0.3 & 11.2 $\pm$ 1.0 & --- & 7.5 $\pm$ 0.1 & 9.8 $\pm$ 0.5 & --- & 6.4 $\pm$ 0.2 & 7.9 $\pm$ 1.0 \\ 
 35.6 $\pm$ 1.7 & --- & 9.8 $\pm$ 0.1 & 13.4 $\pm$ 0.8 & --- & 7.7 $\pm$ 0.0 & 8.1 $\pm$ 0.8 & --- & 7.4 $\pm$ 0.2 & 7.5 $\pm$ 0.0 \\ 
 39.3 $\pm$ 2.0 & 0.3 $\pm$ 0.1 & 11.2 $\pm$ 0.1 & 9.1 $\pm$ 1.5 & 0.3 $\pm$ 0.1 & 8.7 $\pm$ 0.0 & 8.8 $\pm$ 0.2 & 0.3 $\pm$ 0.1 & 8.3 $\pm$ 0.2 & 7.4 $\pm$ 0.1 \\ 
 43.6 $\pm$ 2.4 & 5.8 $\pm$ 0.1 & 11.5 $\pm$ 0.1 & 8.8 $\pm$ 0.7 & 5.4 $\pm$ 0.1 & 9.3 $\pm$ 0.0 & 7.9 $\pm$ 0.7 & 4.4 $\pm$ 0.1 & 8.8 $\pm$ 0.2 & 8.2 $\pm$ 0.3 \\ 
 48.8 $\pm$ 2.8 & 10.1 $\pm$ 0.1 & 10.8 $\pm$ 0.1 & 6.8 $\pm$ 1.3 & 8.7 $\pm$ 0.0 & 10.3 $\pm$ 0.0 & 7.5 $\pm$ 0.8 & 7.2 $\pm$ 0.1 & 9.4 $\pm$ 0.1 & 6.8 $\pm$ 0.2 \\ 
 54.9 $\pm$ 3.4 & 11.0 $\pm$ 0.1 & 10.7 $\pm$ 0.1 & 4.5 $\pm$ 1.5 & 9.3 $\pm$ 0.1 & 9.8 $\pm$ 0.1 & 7.3 $\pm$ 0.2 & 8.5 $\pm$ 0.1 & 9.5 $\pm$ 0.1 & 7.7 $\pm$ 1.4 \\ 
 62.3 $\pm$ 4.1 & 13.3 $\pm$ 0.0 & 9.4 $\pm$ 0.2 & 4.7 $\pm$ 0.9 & 11.1 $\pm$ 0.1 & 10.4 $\pm$ 0.1 & 6.0 $\pm$ 0.2 & 9.9 $\pm$ 0.1 & 9.9 $\pm$ 0.0 & 6.7 $\pm$ 0.5 \\ 
 71.4 $\pm$ 5.1 & 13.2 $\pm$ 0.1 & 7.5 $\pm$ 0.3 & 1.5 $\pm$ 0.8 & 12.1 $\pm$ 0.1 & 9.4 $\pm$ 0.0 & 5.0 $\pm$ 0.1 & 11.2 $\pm$ 0.1 & 9.6 $\pm$ 0.0 & 7.1 $\pm$ 0.2 \\ 
 82.8 $\pm$ 6.4 & 13.2 $\pm$ 0.1 & 5.1 $\pm$ 0.2 & 0.7 $\pm$ 0.4 & 12.2 $\pm$ 0.1 & 7.0 $\pm$ 0.1 & 3.5 $\pm$ 0.8 & 11.5 $\pm$ 0.1 & 8.5 $\pm$ 0.2 & 4.8 $\pm$ 2.5 \\ 
 97.2 $\pm$ 8.2 & 12.1 $\pm$ 0.1 & 3.1 $\pm$ 0.1 & 0.4 $\pm$ 0.2 & 12.6 $\pm$ 0.1 & 6.1 $\pm$ 0.1 & 0.9 $\pm$ 0.5 & 12.4 $\pm$ 0.1 & 7.1 $\pm$ 0.4 & 2.9 $\pm$ 1.5 \\ 
 116.0 $\pm$ 10.9 & 9.4 $\pm$ 0.1 & 1.4 $\pm$ 0.1 & --- & 10.9 $\pm$ 0.1 & 3.0 $\pm$ 0.1 & 0.2 $\pm$ 0.1 & 11.8 $\pm$ 0.1 & 4.2 $\pm$ 0.4 & 1.2 $\pm$ 0.6 \\ 
 141.3 $\pm$ 14.9 & 6.8 $\pm$ 0.1 & 0.4 $\pm$ 0.0 & --- & 9.3 $\pm$ 0.1 & 1.6 $\pm$ 0.1 & 0.1 $\pm$ 0.0 & 10.2 $\pm$ 0.1 & 2.8 $\pm$ 0.2 & 0.1 $\pm$ 0.1 \\ 
 176.9 $\pm$ 21.5 & 3.4 $\pm$ 0.1 & --- & --- & 5.4 $\pm$ 0.1 & 0.3 $\pm$ 0.0 & --- & 7.6 $\pm$ 0.1 & 1.0 $\pm$ 0.0 & --- \\ 
 229.8 $\pm$ 33.2 & 1.2 $\pm$ 0.0 & --- & --- & 2.1 $\pm$ 0.1 & --- & --- & 3.7 $\pm$ 0.1 & 0.1 $\pm$ 0.0 & --- \\ 
 315.5 $\pm$ 57.3 & 0.2 $\pm$ 0.0 & --- & --- & 0.5 $\pm$ 0.0 & --- & --- & 1.0 $\pm$ 0.0 & --- & --- \\ 
\hline
\end{tabular}
}
\label{tab:yield_vs_energy_30deg}
\end{table}

\end{document}